\documentclass[aps,pre,twocolumn,showpacs,preprintnumbers,amsmath,amssymb,superscriptaddress]{revtex4}
\usepackage{graphicx}
\usepackage{bm}
\usepackage{psfrag}
\usepackage{subfigure}

\newcommand{\be}{\begin{equation}}
\newcommand{\ee}{\end{equation}}
\newcommand{\ba}{\begin{eqnarray}}
\newcommand{\ea}{\end{eqnarray}}

\newcommand{\fm}{\langle f \rangle}

\newcommand{\omegaBP}{\omega_{\scriptscriptstyle\rm BP}}

\begin{document} 
\author{Carolina Brito}
\affiliation{Instituto de F\'{\i}sica,  Universidade Federal do Rio
        Grande do Sul, Porto Alegre, Brazil}%


\author{Matthieu Wyart}
\affiliation{Division of Engineering and Applied Sciences, Harvard University,
  Pierce Hall, 29 Oxford Street, Cambridge, Massachusetts 02198, USA}

\date{\today}

\title{Normal modes analysis of the microscopic dynamics in hard discs}


\begin{abstract}
We estimate numerically the normal modes of the free energy in a glass of hard discs. We observe that, near the glass transition or after a rapid quench deep in the glass phase, the density of states  (i) is characteristic of a marginally stable structure,   in particular it displays a  frequency scale $\omega^*\sim p^{1/2}$, where $p$ is the pressure and (ii)  gives a faithful representation of the short-time dynamics. This brings further evidences  that the boson peak near the glass transition corresponds to the relaxation of marginal modes of a weakly-coordinated structure, and implies that the mean square displacement in the glass phase is anomalously large and  goes as $\langle \delta R^2\rangle\sim p^{-3/2}$,  a prediction that we check numerically. 
\end{abstract}

\pacs{61.43.-j, 64.70.Pf, 67.55.Jd}
\maketitle

There is still no accepted description of why the dynamics of liquids rapidly slows down as the temperature is lowered \cite{tarjus}.
Goldstein \cite{goldstein} proposed that the glass transition is related to a change in the topology of the free energy landscape. In this view,
below some temperature meta-stable states appear, the dynamics becomes activated and rapidly slows down. A similar scenario occurs in mean field glass models \cite{wolynes, parisi,laloux}, 
which share formal similarity with the Mode Coupling Theory of liquids \cite{sjorgen}.  The presence of an elastic instability, corresponding to the onset of activation,  is also predicted in systems were particles are deposited randomly (corresponding to an infinite temperature), as the density is changed \cite{argentin}. Nevertheless, the real-space interpretation of those views is not clear. Empirically, the presence of an elastic instability near the glass transition is suggested by Raman and neutron spectra,  which display a peak, the ``Boson Peak", typically shifting toward zero-frequency as the glass is heated toward the glass transition\cite{tao}. This observation is also supported by numerical simulations \cite{argentin2}. The nature of the modes  forming the peak and  how they depend on the structure of the glass are long-standing questions \cite{angell}, on which recent progress was made in a-thermal  assemblies of particles \cite{matthieu1,ning}. Elucidating these questions at  finite temperature, in particular in the vicinity of the glass transition, is necessary to make progress in our microscopic understanding of the glass transition. Indeed these modes, which characterize the short-term dynamics, presumably affect how glasses flow on much longer time-scales, as supported by the observations that the short term dynamics (i) correlate with the fragility of the glass \cite{Sokolov} and (ii) is a good local predictor of the propensity (the tendency of a region  to flow) \cite{harowell}.  To study these modes,  various numerical technics can be used, such as the instantaneous normal modes  analysis \cite{keyes} or the analysis of the vibrational spectrum of inherent structures \cite{saddle}. Nevertheless, these approaches consider the energy rather than the free energy landscape. Such analysis are not adapted for colloidal systems or shaken grains \cite{durian, dauchot} 
or when non-linearities of the interacting potential are important, which is presumably often the case near the glass transition \cite{chandler}.

In this Letter, we study the microscopic dynamics of hard discs in the vicinity of the glass transition and deep in glass phase, via an analysis of the free energy landscape.
To achieve it, we use a recent technic developed to infer the normal
modes of the free energy \cite{brito}.
We show that, (i) after a rapid quench in the glass phase or near the glass transition, the density of normal modes is characteristic of a marginally rigid solid, in agreement with a former analysis of the microscopic structure of the glass \cite{brito}, (ii) throughout the spectrum, the computed frequency of a mode gives the correct estimate for the characteristic time and amplitude of its fluctuations. This demonstrates that our mode analysis performs well in the vicinity of the glass transition. (iii)  These results justify the presence of a boson peak and an anomalously large  and slow  microscopic relaxation. In particular, the mean square displacement must vary as $\langle \delta R^2\rangle\sim p^{-3/2}$, in good agreement with our simulations. This signifies that, deep in the glass phase, the amplitude of the particles  trajectories become infinitely larger than the instantaneous size of the cages.

As the viscosity of a super-cooled liquid increases, the dynamics become intermittent \cite{kob,heuer,brito2}: there are long quiet periods where particles are rattling around their average positions, interrupted
by rapid rearrangements toward new configurations. In a simulation, this intermittent dynamics can be followed via the  self density  correlation function $C(\vec q, t, t_w) = \langle e^{ i \vec q . (\vec R_j(t+t_w) - \vec R_j(t_w) )} \rangle_j$, where the average is made on all particles, but not on time, and where  $\vec R_j(t)$ indicates the position of particle $j$ at time $t$. This observable displays long plateaus, interrupted by sudden jumps leading to new plateaus \cite{brito,brito2}. We shall refer to these plateaus as meta-stable states. In the reference \cite{brito2}, the transition between states was studied. Here we focus on their inner dynamics. 

As shown in \cite{brito}, in meta-stable states the  Gibbs free energy of a hard sphere system can be approximated as:
\be
\label{1}
{\cal G}= - k_bT \sum_{ ij} \ln \langle h_{ij}\rangle,
\ee
where the sum is made on all pairs of particles $ij$ {\it in contact},
i.e colliding with each other in a specific meta-stable state.
$T$ is the temperature,  
$\langle h_{ij}\rangle\equiv \langle ||{\vec R_i}-{\vec
  R_j}||\rangle-(\sigma_i+\sigma_j)/2$ is the average gap between
particles  measured in a given meta-stable state,
and $\sigma_i$ is the diameter of particle $i$. Eq.(\ref{1})
draws an analogy between the free energy of hard spheres and the
energy of a system of logarithmic springs. Eq.(\ref{1}) can be expanded
for small displacements around the average particles position
$\delta\vec{R_i} = \vec R_i - \langle \vec R_i\rangle$. This enables
to define the dynamical matrix  ${\cal M}$ \cite{ashcroft}, which
describes how the average particles positions respond to a small
external force.  The eigenvectors $|\delta R^\omega\rangle$ of ${\cal
  M}$ are the vibrational modes \cite{clapack}. For particles of unit mass, we define as usual their frequency $\omega$ as the square  root of the corresponding eigenvalue of  ${\cal M}$. 
 In what follows $\omega$ is used to label the normal modes. We shall see below that this label corresponds, within a good approximation, to the physical oscillation frequency of the mode.

In practice, we use an event-driven code to simulate a two-dimensional bidisperse 
hard discs system \cite{brito,brito2} and   compute ${\cal M}$. 
Half of the particles have  a unit diameter, while the others 
a diameter of  1.4 and their mass is $m\equiv1$. 
 We choose $k_bT$ as our unit of energy. 
We can equilibrate this system up to $\phi<\phi_0\approx 0.79$. To  generate larger packing fractions and enter in the aging regime, we proceed as follows: we decompress 
 jammed configurations of \cite{J} of packing fraction  $\phi_c\approx 0.83$ by decreasing the particle diameters by some fraction, and  assign a random velocity to every particle. 
As the dynamics run, we compute the  self-density correlation function and determine its plateaus or meta-stable states. For each of them, we consider a time interval $[t,t+t_1]$ corresponding to a single state,
 where $t_1$ is chosen to be much larger than the average collisional time of a contact $\tau_c$.
 In what follows $t_1$ corresponds typically to 100 $\tau_c$ but our results are robust to other choices as long as $t_1\gg\tau_c$.
 The network of contacts and the average particle positions $\langle \vec R_i\rangle$ are computed during this interval. This enables to compute ${\cal G}$ and ${\cal M}$ numerically.

\begin{figure}                                          
   \rotatebox{-90}{\resizebox{5.8cm}{!}{\includegraphics{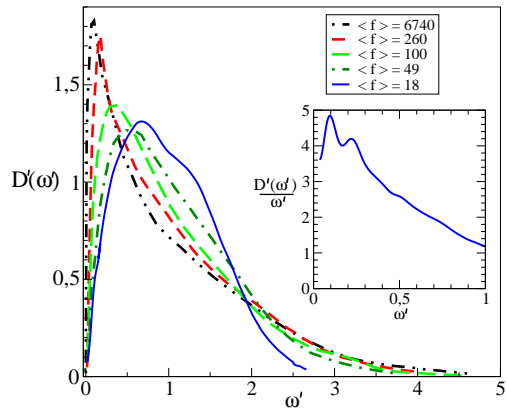}}}
   \vspace{-0.3cm}
   \caption{Densities of states  $D'(\omega')\equiv \fm D(\omega)$  {\it vs.} rescaled frequency
     $\omega'=\omega/\langle f\rangle$ for   different values of  $\langle f\rangle$ in a system of 
     $N=256$ particles.      We used $t_1 = 5\times 10^4$. Inset: $D(\omega')/\omega'$ {\it vs.} $\omega'$  for $\langle f\rangle=18$. }
   \label{Dw_examples}
 \end{figure}
 
\begin{figure}
  \rotatebox{-90}{\resizebox{5.7cm}{!}{\includegraphics{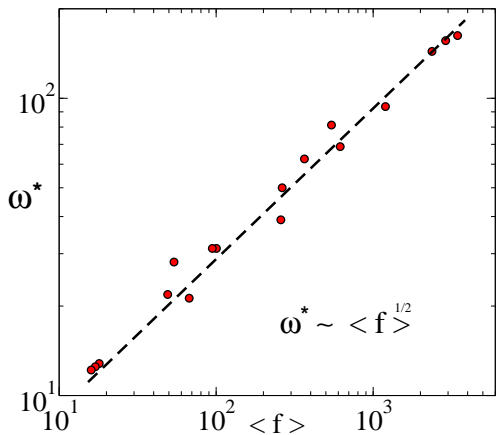}}}
\vspace{-0.3cm}
  \caption{Characteristic frequency $\omega^*$ as defined in the text  {\it vs}  average force $\fm$. 
     Slashed line corresponds to the fit $\omega^*\sim \langle f\rangle^{1/2}$.}
  \label{wstar_vs_f}
\end{figure}

To compute the normal modes and the density of states,
 we diagonalize ${\cal M}$.
 Results are shown in Fig.(\ref{Dw_examples}).  
Curves are labeled by the average contact force
 $\langle f\rangle\equiv\langle 1/\langle h_{ij}\rangle \rangle_{ij}$, 
where the average is made on all contacts.
 $\langle f \rangle$ diverges near $\phi_c$ and scales as the pressure.
 The densest 
system we can equilibrate corresponds to  $\langle f \rangle\approx 18$. 
For larger  $\langle f \rangle$ 
the system is a glass. To compare various packing fractions, the frequency 
axis is rescaled by the square root of the typical interaction stiffness $k$,
 of order $k^{1/2}\sim \langle f\rangle$ in our units. 
We observe that: (i) for all $\langle f\rangle$, $D(\omega)$ 
increases rapidly from zero-frequency to reach a maximum at some frequency $\omega^*$, before
 decaying again. In our finite size system the modes constituting the low-frequency part of the spectrum are not plane-wave like \footnote{For examples of 
lowest-frequency modes, see \cite{brito2}.}. This can be inferred  from the inset of  Fig.(\ref{Dw_examples}), where $D(\omega)$ is normalized by its Debye value ($\sim \omega$ in two dimensions): no plateau can be detected at low frequency, where a peak appears at some frequency $\omegaBP$ significantly smaller than  $\omega^*$.
(ii) $\omega^*$ decreases 
toward zero in the rescaled units as $\langle f\rangle$ diverges.
Fig.(\ref{wstar_vs_f}) shows the dependence of  $\omega^*$ with $\langle f\rangle$. We observe that the scaling
 \be
 \label{2}
 \omega^*\sim \langle f\rangle^{1/2}
 \ee
 holds well from the glass transition toward our densest packing. Scaling behaviors in the vibrational spectrum, together with the fact that the lowest-frequency vibrations observed are not plane wave like, were first observed in  elastic non-thermal particles \cite{J}.
 This was later explained as follows: in a weakly-coordinated network of springs at rest, i.e. with a small coordination number $z$ close to a critical value $z_c=2d$ ($d$ is the spatial dimension),
 extended modes, quite different from plane waves, appear already in the low-frequency part of the spectrum. The frequency scale of these {\it anomalous modes} is $\omega^*\sim k^{1/2} (z-z_c)$
  \cite{matthieu1}.
 In a system of repulsive particles with non-zero forces, these modes are shifted toward  a lower frequency \cite{matthieu2}.
   To avoid an elastic instability, 
 the system must display enough contacts. In the marginally stable case where there are just enough contacts  to be rigid, the spectrum displays anomalous modes
 up to zero frequency, but present a characteristic frequency at $\omega^*\sim \langle f\rangle^{1/2}$, such that $D(\omega^*)$ is of order of its typical value $k^{-1/2}$. This estimate of $D(\omega^*)$  agrees with Fig.(\ref{Dw_examples}), where $D(\omega^*)$  appears to remains of the same order in rescaled frequency.  Thus both the scaling of $\omega^*$ and the observation that anomalous modes are present at frequency $\omegaBP<<\omega^*$  support that the glass phase lies close to marginal stability, as was already suggested by a direct measure of the coordination in the glass phase \cite{brito}.

\begin{figure}                                
   \rotatebox{-90}{\resizebox{5.9cm}{!}{\includegraphics{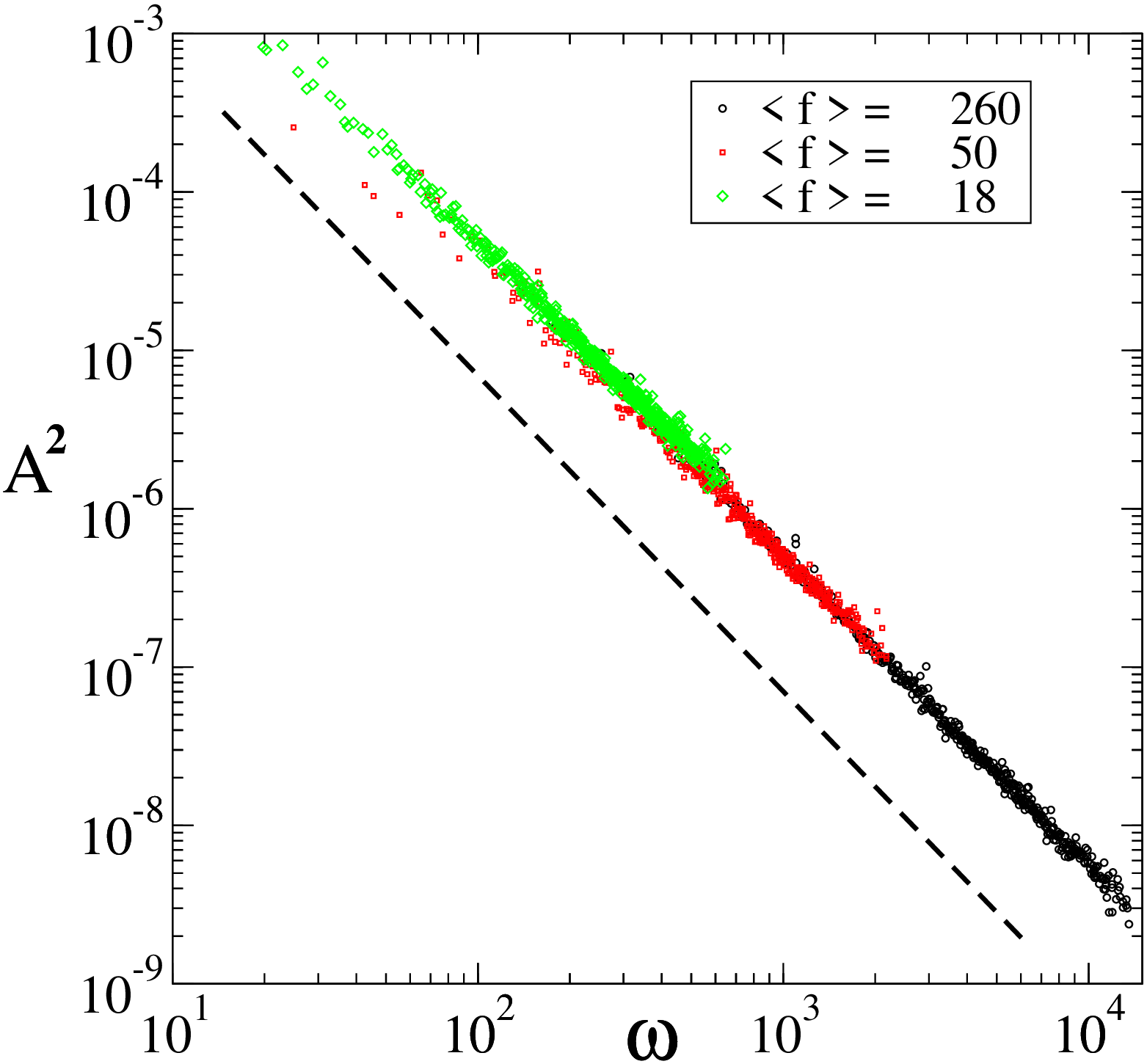}}}
   \vspace{-0.2cm}
   \rotatebox{-90}{\resizebox{5.9cm}{!}{\includegraphics{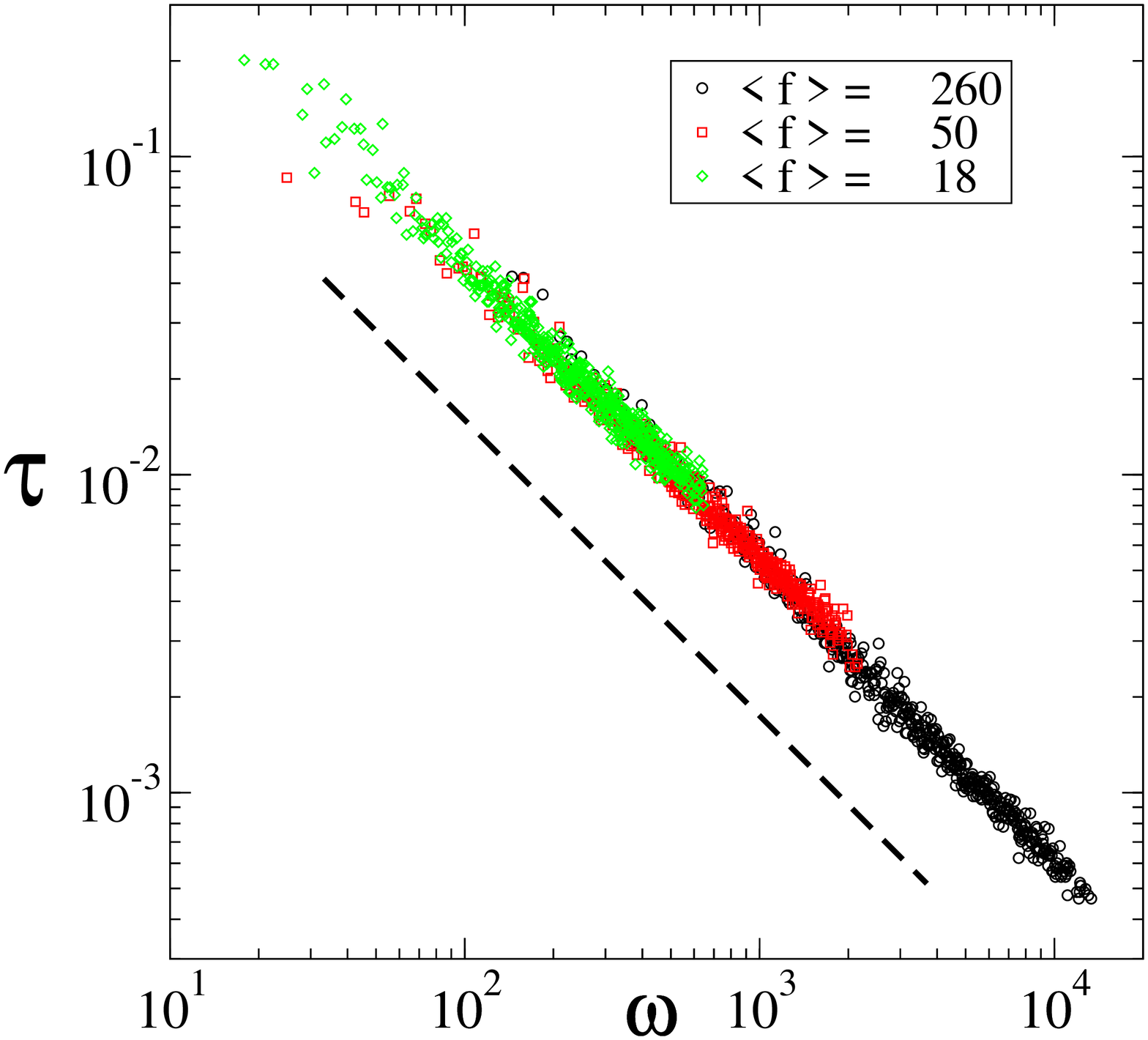}}}
   \caption{(a) Average squared amplitude of the modes $\langle A^2(\omega)\rangle $ {\it vs} $\omega$ at various packing fractions in a system of 256 particles, 
     both in the glass phase ($\fm=260$ and $\fm=50$) or in the 
     super-cooled liquid ($\fm=18$).
    Each point corresponds to one mode. The dashed line corresponds to the 
    fit $\langle A^2(\omega)\rangle\sim 1/\omega^2$.
    (b) Relaxation time $\tau(\omega)$ of each mode {\it vs} $\omega$ for the same packing fraction.
  The slashed line corresponds to the relation $\tau(\omega) \sim 1/\omega$.}
    \label{A_vs_w}
\end{figure}

We check that  $D(\omega)$ reflects well the dynamics.
We  project the dynamics on the normal modes and define for each frequency $\omega$:
\ba
C_{\omega} (t) =  \langle~ \langle \delta R(t+t_w) | \delta R^{\omega} \rangle ~.~
  \langle \delta  R(t_w) |  \delta R^{\omega}  \rangle   ~\rangle_{t_w},  
\label{3}
\ea 
  where 
$| \delta R(t+t_w) \rangle$ is the displacement field from the average configuration, $\langle \delta R^a | \delta R^b \rangle\equiv \sum_i \delta {\vec R^a}_i\cdot  \delta {\vec R^b}_i$, and where the average is made on all time segments $[t_w,t_w+t]$  included in a meta-stable state. 
If the modes were longitudinal plane-waves, $C_{\omega} (t)$ would describe the relaxation of density fluctuations. Here this is not at all the case.
We compute the amplitude $A(\omega)$ of a mode fluctuation defined as  $\langle A^2(\omega)\rangle =C_{\omega}(0)$, and a characteristic time $\tau(\omega)$ of the mode relaxation,
defined as the earliest time at which $C_{\omega}(t)$ has decayed by 10\%:  $C_\omega(\tau(\omega)) = 0.9 C_{\omega}(0)$. Taking other thresholds (30\% or 50\%) leads to the same scaling result but reduce our statistics. 
The dependence of these quantities with frequency are respectively shown in  Fig.(\ref{A_vs_w}-a) 
and Fig.(\ref{A_vs_w}-b) for three meta-stable states at different pressure.
 Two configurations are in the glass phase and one in the super-cooled liquid phase. 
In all cases, these quantities were computed for each mode of the spectrum.
We observe that the modes display weakly damped oscillations, whose amplitude and period follow:
 \ba
 \langle A^2(\omega)\rangle\sim \frac{1}{\omega^2}, \\
\label{rtt1}
 \tau(\omega) \sim \frac{1}{\omega}.
 \label{rtt}
 \ea
These results hold true even for the low-frequency part of the
spectrum, although  more scattering is found there \footnote[3]{Sometimes one or a few unstable modes are observed at very low frequency, indicating a shoulder or a saddle in the free energy landscape. In this case the values of $\langle A^2 \rangle$ and $\tau$ are found to be of the order of those of the lowest-frequency stable modes.}.
Thus, even for the lowest-frequency part of the spectrum, the  non-linearities do not appear to dominate the mode dynamics in a dense disordered assembly of colliding hard spheres, although the interacting potential is discontinuous.

This observation implies that $D(\omega)$  gives a rather faithful distribution of relaxation time scales of the microscopic dynamics. This allows us to identify the peak apparent in the inset of Fig.(\ref{Dw_examples}) as the Boson Peak, which appears as a similar hump in Raman or neutron spectra in molecular liquids \cite{tao, angell}.  Near the glass transition, this peak  appears at a frequency significantly smaller than $\omega^*$. The scaling behavior of $D(\omega)$ supports that these modes correspond the anomalous modes characterizing  weakly-coordinated structures, whose geometry is studied in \cite{matthieu2}.  It would be interesting to go further and test if the early-$\beta$ relaxation \cite{sjorgen}, which characterizes the relaxation on time scales longer than the boson peak,  is due to the weakly-damped relaxation of anomalous modes even softer than the peak, or to other non-linear processes not captured by the present analysis. Unfortunately this relaxation is difficult to study numerically using Newtonian dynamics, and our present analysis should be performed in three-dimensions, where numerical and experimental data exist, to test this possibility.

In what follows we rather focus on the amplitude of the short-time relaxation, in particular the  particle mean-square displacements $\langle \delta {\vec R}_i^2 \rangle$. 
Assuming that the dynamics of different modes is independent, one gets:
\be
\label{21}
\langle {\delta \vec R_i}^2\rangle= \sum_\omega \langle A^2(\omega)\rangle {\delta \vec R_i^\omega}^2
\ee

We then average on all particles and define $\langle\delta \vec R^2 \rangle =  1/N~\sum_i \langle {\delta \vec R_i}^2\rangle$ where $N$ is the system size. Using the modes normalization  $\langle {\delta \vec R_i(\omega)}^2\rangle_i = 1/N$ and applying  Eqs.(\ref{rtt1}) lead to:
\be
\label{22}
 \langle {\delta \vec R}^2\rangle\sim \int_0 \frac{D(\omega)}{\omega^2} d\omega\geq \int_{\omega^*}\frac{D(\omega)}{\omega^2} d\omega
\ee
Since $D(\omega)$ reaches the order of its typical value $1/\sqrt
k\sim 1/\langle f\rangle$ for $\omega\geq \omega^*$, the last integral is dominated by the lowest bound and one gets:
\be
\label{R2}
\langle {\delta \vec R}^2\rangle \geq
\frac{D(\omega^*)}{\omega^*}\sim \fm ^{-3/2}
\ee

It is easy to show from Eq.(\ref{22}) that the scaling (\ref{R2})
becomes an equality if $D(\omega)$ does not increase more rapidly than
linearly between 0 and $\omega^*$ \footnote[2]{In two dimensions, this analysis is only possible in a finite size system, since plane waves lead to a log $N$ divergence of the mean square displacement.}. 
 The bound (\ref{R2}) holds in any dimension $d\geq 2$, as long as the marginal stability of the structure and the harmonic behavior of the modes is assumed.  Eq.(\ref{R2}) must be
compared with the fluctuations of a crystal:  $\langle \delta{\vec R_i}^2\rangle \sim h^{2} \sim \fm ^{-2} $ (with $\log N$ corrections in two dimensions). In other words,  near maximum packing, the amplitude of 
particles motions becomes infinitely  smaller in the crystal, where it is simply of the order 
of the inter-particle gap $h$.  In the hard sphere glass,  particles diffuse much more than the size of their 
instantaneous cage $h$, and the size of the region they visit in a meta-stable state is at least of order
 $h^{3/4}$. 

\begin{figure}
\begin{center}                             
  \rotatebox{-90}{\resizebox{5.8cm}{!}{\includegraphics{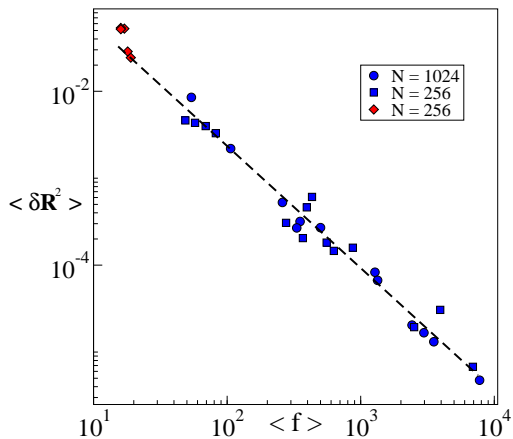}}}
\vspace{-0.2cm}
  \caption{Mean square displacement $\langle \delta \vec R^2 \rangle$ versus
average contact force $\fm$ for  $N=1024$ (circles) and $N=256$
(squares) particles. Diamonds (red points) correspond to equilibrated systems of  $N=256$ 
particle.
Slashed line corresponds to the best fit of the form
$\langle \delta \vec R^2 \rangle \sim \fm^{-3/2}$.}
  \label{sqrt_r2} 
  \end{center}                             
\end{figure}

To check  this prediction, we consider various meta-stables states, 
and measure the mean square displacement $\langle \delta \vec R^2 \rangle$ by averaging on
 some long time  $t_1$, see  Fig.(\ref{sqrt_r2}). 
Our numerical result agrees well with (\ref{R2}). As for Eq.(\ref{2}), the scaling prediction 
captures the behavior also near the glass transition,  where the system can be equilibrated.

We have used and tested a free energy-based approach that allows us to perform a mode analysis at non zero-temperature, in particular in the vicinity of the glass transition. We have observed that, near the glass transition or after a rapid quench, the spectrum of the free energy is characteristic of a marginally rigid solid, as confirmed further by the scaling of the amplitude of the mean square displacement in the glass phase. The observation that on the time scales we can probe the system remains near the onset of rigidity supports the proposition, expressed originally by Goldstein in terms of free energy landscape, that the dynamics dramatically slows down once meta-stable states appear.  Our analysis provides a spatial interpretation of this landscape picture, where the elastic instability is governed at a microscopic level by coordination and contact forces, and where the soft modes at play are collective rearrangements, whose characteristic length scale is limited near the glass transition and diverges near the maximum packing of hard spheres \cite{matthieu1}. Finally, the observation that  the microscopic dynamics, and activation events \cite{brito2}, are dominated by soft modes is consistent with the observation that the short term dynamics is a good indicator of propensity \cite{harowell}. It suggests that correlations between coordination, contact force, and propensity may exist; and that soft modes may be the elementary objects to consider to describe microscopically activation.

\begin{acknowledgments}
We thank G. Biroli, J-P. Bouchaud, L. G. Brunnet,   D. Fisher,   S. Nagel
and T. Witten for helpful discussion and L.Silbert for furnishing the initial 
jammed configurations. C. B. was supported by CNPq and  M. W. by the
George Carrier Fellowship.
\end{acknowledgments}

\end{document}